\pdfoutput=1 
\documentclass[sigplan,screen,nonacm]{acmart}
\renewcommand\footnotetextcopyrightpermission[1]{}
\makeatletter
\renewcommand*{\NAT@spacechar}{~}
\makeatother

\usepackage{csquotes}
\usepackage{balance}
\usepackage{xcolor}
\usepackage[normalem]{ulem}

\newcommand{\code}[1]{\texttt{#1}}

\newif\ifshowtodos
\showtodostrue %

\newif\ifshowdiscuss
\showdiscusstrue %

\usepackage{tikz}
\makeatletter
\newenvironment{btHighlight}[1][]
{\begingroup\tikzset{bt@Highlight@par/.style={#1}}\begin{lrbox}{\@tempboxa}}
{\end{lrbox}\bt@HL@box[bt@Highlight@par]{\@tempboxa}\endgroup}

\newcommand\btHL[1][]{%
  \begin{btHighlight}[#1]\bgroup\aftergroup\bt@HL@endenv%
}
\def\bt@HL@endenv{%
  \end{btHighlight}%
  \egroup
}
\newcommand{\bt@HL@box}[2][]{%
  \tikz[#1]{%
    \pgfpathrectangle{\pgfpoint{1pt}{0pt}}{\pgfpoint{\wd #2}{\ht #2}}%
    \pgfusepath{use as bounding box}%
    \node[anchor=base west, fill=codehighlightgreen!30,outer sep=0pt,inner xsep=1pt, inner ysep=0pt, rounded corners=3pt, minimum height=\ht\strutbox+1pt,#1]{\raisebox{1pt}{\strut}\strut\usebox{#2}};
  }%
}
\makeatother

\usepackage{listings}
\usepackage{lore}
\usepackage{color}
\usepackage{listings}

\definecolor{codehighlightgreen}{HTML}{81C784}
\definecolor{codegreen}{HTML}{008000}
\definecolor{codegray}{HTML}{606060}
\definecolor{codeorange}{HTML}{FFA040}
\definecolor{codered}{HTML}{D04020}
\definecolor{codeblue}{HTML}{06287e}
\definecolor{codelightblue}{HTML}{0e84b5}
\definecolor{codepurple}{HTML}{B00060}
\definecolor{comment}{HTML}{2d2d2d}
\definecolor{light-gray}{gray}{0.92}
\definecolor{gray}{rgb}{0.5,0.5,0.5}
\lstdefinestyle{scitzenCodestyle}{
  frame=none,
  showstringspaces=false,
  basicstyle={\small\ttfamily},
  keywordstyle={\small\ttfamily},
  numbers=left,
  xleftmargin=2.3em,
  numberstyle=\fontsize{7}{9}\selectfont\color{gray}\bfseries,
  breaklines=true,
  breakatwhitespace=true,
  tabsize=2,
  escapeinside={(*@}{@*)},
  numberblanklines=true,
  firstnumber=last,
  literate={"}{\textquotedbl}1,
  captionpos=b
}
\lstdefinestyle{highlightedCodestyle}{
  language=lore,
  frame=none,
  showstringspaces=false,
  basicstyle={\small\ttfamily},
  keywordstyle={\bfseries},
  keywordstyle=[2]{},
  numbers=left,
  xleftmargin=2.3em,
  numberstyle=\fontsize{7}{9}\selectfont\color{gray}\bfseries,
  breaklines=true,
  breakatwhitespace=true,
  tabsize=2,
  numberblanklines=true,
  firstnumber=last,
  literate={"}{\textquotedbl}1,
  captionpos=b,
  abovecaptionskip=5pt,
  moredelim=**[is][\btHL]{@}{@}
}
\begin{document}

\title{Distributed Locking as a Data Type}

\author{Julian Haas}
\orcid{0000-0001-9959-5099}
\affiliation{%
    \institution{Technische Universität Darmstadt}
    \country{Germany}
}

\author{Ragnar Mogk}
\orcid{0000-0003-4583-1791}
\affiliation{%
    \institution{Technische Universität Darmstadt}
    \country{Germany}
}

\author{Annette Bieniusa}
\orcid{0000-0002-1654-6118}
\affiliation{%
    \institution{University of Kaiserslautern-Landau}
    \country{Germany}
}

\author{Mira Mezini}
\orcid{0000-0001-6563-7537}
\affiliation{%
    \institution{Technische Universität Darmstadt}
    \country{Germany}
}

\begin{abstract}
  Mixed-consistency programming models assist programmers in designing applications that provide high availability while still ensuring application-specific safety invariants.
However, existing models often make specific system assumptions, such as building on a particular database system or having baked-in coordination strategies.
This makes it difficult to apply these strategies in diverse settings, ranging from client/server to ad-hoc peer-to-peer networks.

This work proposes a new strategy for building programmable coordination mechanisms based on the \emph{algebraic replicated data types} (ARDTs) approach.
ARDTs allow for simple and composable implementations of various protocols, while making minimal assumptions about the network environment.
As a case study, two different locking protocols are presented, both implemented as ARDTs.
In addition, we elaborate on our ongoing efforts to integrate the approach into the \emph{LoRe} mixed-consistency programming language.

\end{abstract}

\maketitle

\section{Introduction}

Nowadays, most software is distributed in one way or the other to provide features such as collaboration, multi-device synchronization, and access to remote information.
However, it is important to ensure that such applications do not impede the \enquote{local} functionalities due to partial failures of remote systems.
Therefore, \citet{kleppmann2019} propose the concept of \emph{local-first software} and outline a set of ideals that such software should follow.
Operationally, these ideals imply that local-first software
operates under weaker consistency models such as eventual or causal consistency.
This ensures offline availability of all operations, regardless of network connectivity.

Local-first is suitable for many collaboration use cases.
However, in large applications dealing with various concerns, stronger guarantees
may be necessary \emph{for certain parts of the software}.
For instance, in a company-wide project management system, planning may work local-first, while budget requests must be enforced globally to maintain the \emph{invariant} that total expenses remain below a certain threshold.

In this type of setting, it is difficult to find a consistency model that suits the entire application.
Using strong consistency models, which involve coordination, sacrifices offline availability and conflicts with the objectives of local-first applications. At the same time, weaker models often cannot guarantee the application's safety requirements.
The best option is to choose a design somewhere in the middle.
However, how should developers determine the appropriate consistency model for each part of their application?
In general, combining different consistency levels is a process that is error-prone and requires expertise in
both, the application domain and distributed systems in general.

Dedicated mixed-consistency programming models such as LoRe~\cite{haasLoReProgrammingModel2024}, ConSyst~\cite{kohlerRethinkingSafeConsistency2020}, Blazes~\cite{Alvaro2014}, and Carol~\cite{Lewchenko2019} provide a solution:
They offer built-in support for mixed-consistency applications
through special programming abstractions and automated reasoning.
They -- and hybrid consistency systems~\cite{Balegas2018, Gotsman2016, Li2012} -- typically select between two \enquote{modes} of operation: An eventually consistent mode that allows concurrent execution, and a strongly consistent mode,
where potentially conflicting operations are coordinated.

These systems often employ a specific baked-in technique to coordinate operations.
For example, 
some~\cite{Alvaro2014,kohlerRethinkingSafeConsistency2020} employ a centralized lock manager such as \emph{Zookeeper}~\cite{huntZooKeeperWaitfree2010}, while others~\cite{haasLoReProgrammingModel2024,Lewchenko2019,Li2012,Gotsman2016} rely on a custom token-passing algorithm.
Each coordination technique makes specific system assumptions.
One technique might be better for a large number of participants while
another 
is more suited 
for extended offline periods.
The \enquote{one size fits all} approach of existing mixed-consistency solutions,
makes it hard to write a program once and
then deploy it in different settings such as geo-replicated servers
vs. peer-to-peer networks.

We argue for programmable coordination mechanisms 
and in this work-in-progress report elaborate on our ongoing efforts to enhance
LoRe~\cite{haasLoReProgrammingModel2024} with such mechanisms.
We propose modelling coordination protocols as replicated data types following the \emph{Algebraic Replicated Data Types} (ARDT)~\cite{kuessnerAlgebraicReplicatedData2023} approach.
The ARDT-based approach has several advantages.
First, they facilitate the implementation of 
diverse protocols through reuse and composition of existing components
while posing minimal requirements on the network environment.
Second, they enable mixing and combining protocols in the same application, 
guided by application specific system and deployment assumptions.
In this paper, we focus on mutual-exclusion protocols, but we
believe that the \enquote{protocols as ARDTs} approach is general 
enough to support a wide range of existing coordination protocols and 
simple enough to allow distributed systems experts to implement their own protocols or 
to adapt existing ones.

In Section~\ref{strategy-implementation}, we demonstrate how to implement mutual-exclusion protocols with ARDTs.
We discuss the benefits of the approach in Section~\ref{discussion} and look at ideas for future work
in Section~\ref{outlook}.
Finally, we discuss related work in Section~\ref{related-work}.

\section{Mutual Exclusion with ARDTs}
\label{strategy-implementation}

We provide background details about Algebraic Replicated Data Types (ARDTs)~\cite{kuessnerAlgebraicReplicatedData2023} and then demonstrate how to use them for mutual-exclusion.

\subsection{Algebraic Replicated Data Types}

An ARDT is a (Scala) algebraic data type%
\footnote{Algebraic data types are types formed by composition of other types, most notably product types such as tuples, structs, etc.}%
, additionally equipped with a merge function that has the lattice properties (idempotent, commutative, associative). Specifically, any Scala type like \code{Set[Int]} can be used as long as we can implement an instance of \code{Lattice[Set[Int]]}, where the \code{Lattice} interface provides the mentioned merge function.
Developers may use these types like any other immutable Scala data type.
As expected from algebraic data types, ARDTs compose: We can declare new data structures where each individual component is an ARDT, and in this case, the Scala compiler can even derive a corresponding \code{Lattice} instance automatically.

Since ARDTs are immutable, calling a method such as \code{add} on a \code{Set} does not mutate the underlying data structure itself.
Instead, the call returns a \emph{delta} that represents the effect of the state mutation.
A runtime system using ARDTs can then merge these deltas with the previous value to produce the new current value.
For example, if the current state is \code{Set(1, 2)}, we can add new value by merging \code{Set(3)}.
Note that always merging deltas implies that some standard \code{Set} operations have no effect in the replicated setting; notably, we need additional data to track removals. For sets, the ARDT library provides a suitable \code{DotSet} -- where dots are globally unique timestamps that are used to track removals.

\subsection{Token-Based Mutual Exclusion}
We start with a simple token-based mutual exclusion protocol: each critical operation that needs to be coordinated is assigned a token, and only the replica (represented by a unique \code{Uid}) that currently has the token may execute the critical operation. Replicas announce that they want to obtain the token and the current owner of a token is expected to hand it over eventually.

\paragraph{Implementing the protocol}

The data type definitions of our ARDT-based implementation are shown in Listing~\ref{tokenclass}. The token consists of a statement of ownership and a set of Uids. The ownership is represented by the current owner's Uid along with an epoch. The epoch is used to decide which statement of ownership to keep while merging -- we only ever keep the newest one. This is reflected in the lattice definition for \code{Ownership}; it derives an \code{Ownership} lattice from the natural order of the epoch (represented by a 64-bit integer value). Specifically, we use the lexicographic ordering on the \code{Ownership} type, which sorts by epoch first.
In case the epoch for two values is the same, a fallback based on the ordering of the Uid type is used (which is arbitrary but consistent).
As \code{Token} is simply a product of two other ARDTs (\code{Ownership} and \code{DotSet}),
its \code{Lattice} can be derived automatically -- by merging individual components while preserving the structure.
With this, we have defined the state of the ARDTs and can rely on causally consistent replication.

\begin{lstlisting}[float=*t,
  caption={Code for the Token data types. Important parts are {\btHL highlighted}.},
  label=tokenclass]
case class @Token@(os: Ownership, wants: DotSet[Uid])
case class @Ownership@(epoch: Long, owner: Uid)

given Lattice[Ownership] = Lattice.fromOrdering(Orderings.lexicographic)
given Lattice[Token] = Lattice.derived
\end{lstlisting}

\begin{lstlisting}[
  float=*t,
  caption={Code for the Token protocol. Important parts are {\btHL highlighted}.},
  label=tokenprotocol,
]
case class Token(os: Ownership, wants: DotSet[Uid]):
  val unchanged: Token = Token(Ownership.unchanged, DotSet.unchanged)

  def @isOwner@(using ReplicaId): Boolean = replicaId == os.owner

  def @request@(using ReplicaId, Dots): Token =
    val deltaSet = wants.addElem(replicaId)
    Token(Ownership.unchanged, deltaSet)

  def @release@(using ReplicaId, Dots): Token =
    val deltaSet = wants.removeElem(replicaId)
    Token(Ownership.unchanged, deltaSet)

  def @upkeep@(using ReplicaId): Token =
    if !isOwner then unchanged
    else selectFrom(wants) match
      case None => unchanged
      case Some(nextOwner) =>
        Token(Ownership(os.epoch + 1, nextOwner), DotSet.unchanged)

  def selectFrom(wants: DotSet[Uid])(using ReplicaId) =
    wants.elements.maxOption.filter(id => id != replicaId)
\end{lstlisting}

Next, we discuss the operations on this ARDT that provide mutual-exclusion semantics (Listing~\ref{tokenprotocol}).
The API for the mutual-exclusion schemes
 consists of four methods:
(1) \code{isOwner} is called to check if the executing replica is the current owner and thus may enter the critical section protected by this token;
(2) \code{request} is called when a replica wants to acquire the token;
(3) \code{release} is called when a replica no longer needs the token;
(4) \code{upkeep} is called regularly, specifically after releasing, and whenever the state changes due to updates from the network.

The \code{isOwner} method checks if the local \code{replicaId} matches the owner stored in the current value.
The \code{using} syntax is a Scala feature to pass implicit contexts around by type; we use it here for the local replica ID (accessed with \code{replicaId}) to denote its special status (as opposed to the ID of any other replica).
When the ownership query returns true, we assume that the current replica is the exclusive owner of that token, thus a critical operation protected by the token may now execute safely.

The \code{request} and \code{release} methods add/remove the current replica from the set of \code{wants} to denote interest in ownership.
They \enquote{modify} the current state by returning a delta, i.e., a value of type \code{Token} that describes the change.
To construct the request/release delta, we use the corresponding (delta producing) methods on \code{DotSet} and wrap the resulting delta into a token with \code{unchanged} ownership.

The \code{upkeep} method first checks if the calling replica is the current owner.
If this is the case, we arbitrarily select a \code{nextOwner} from the set of \code{current.wants}.
Otherwise, the \code{unchanged} delta is returned. 
If there is a \code{nextOwner} that was selected, we increase the ownership epoch by one
and return the tuple as a new delta (together with the unchanged/empty set since the set \code{current.wants} does not change in the \code{upkeep} method).

\paragraph{Correctness of the protocol}

For mutual exclusion, we require that \emph{at most one replica believes it is the current owner}.
Intuitively, this is true because the \code{upkeep} method transfers ownership from one replica to another, similar to handing over a physical token.

We reason about ARDTs as sets of deltas and about replicas as subsets of observed deltas. Delayed/dropped deltas are modelled by subsets and duplication/reordering of deltas does not matter because of the lattice properties. For correctness of the mutual exclusion protocols, we additionally require that each delta is produced by one of the operations,
i.e., there are no (undetected) bit flips or incorrect replicas.

We can prove the stated mutual exclusion property by induction on the creation of deltas.
To this end, let us consider ownership deltas -- pairs of an epoch (a Long) and the \emph{target owner} (a Uid) of that epoch.
Our induction invariant is that if there is a delta $d$ with target owner $r$, then either (1) there is no $d'$ with $d < d'$ (i.e., no other delta is greater -- here: from a more recent epoch -- according to the lattice order) or (2) $r$ has observed a $d'$ with $d < d'$.
Intuitively, there exists exactly one greatest delta%
\footnote{Note that the ownership lattice is a total order, which is not true for all lattices.}
for which (1) is true, and thus the target owner has not yet observed a greater delta.
For every other ownership delta, (2) is true because the delta has a corresponding transfer that must be visible to the owner (who did the transfer).
Thus, our invariant implies mutual exclusion: the delta as in part (1) of the invariant has the only potential owner as its target.

We initialize the system with a delta that assigns one of the replica IDs to be the owner in the initial epoch, which trivially fulfills the invariant.
In any system where the invariant is fulfilled, only the replica that has observed the greatest delta $d$ can be the owner, thus produce a (non-trivial) ownership delta $d'$.
Thus, $d'$ is the new greatest delta%
\footnote{Specifically, due to the monotonicity properties of lattices, when operations return smaller or equal deltas (such as all the unchanged deltas) these are immediately discarded as they do not change the current state.}
in the system, and both parts of the invariant are fulfilled%
\footnote{Invariant part (1) is fulfilled for $d'$ because there is no greater delta and part (2) is fulfilled for $d$, because $d'$ is immediately observed by the generating replica $r$.}.

This simple protocol does not guarantee progress under (permanent) node failure. In cases where the transfer of ownership does not happen -- e.g., because the current owner fails permanently -- the token transfer is stuck.

\subsection{Voting-Based Mutual Exclusion}
The second mutual exclusion algorithm is a simple majority-voting approach assuming a fixed number of participants.
\begin{lstlisting}[
  float=*t,
  caption={Data type definitions for the voting protocol.},
  label=votingprotocol-datatypes
]
case class Vote(owner: Uid, voter: Uid)
case class Epoch[E](counter: Time, value: E)
case class Voting(rounds: Epoch[DotSet[Vote]])

given Lattice[Voting] = Lattice.derived
\end{lstlisting}
Listing~\ref{votingprotocol-datatypes} shows the state definition of the \code{Voting} ARDT. 
We support multiple rounds of voting, represented as the \code{Epoch} type. 
It works similarly to the \code{Ownership} ARDT, except that when the epoch counter of two deltas are equal, 
we merge the value stored in the epoch. This allows us to contain a set of votes within the epoch.
We use the same interface of four operations for mutual exclusion.
We only give a high-level overview of their behavior below -- the full source code can be found in the Appendix in Listing~\ref{votingprotocol}.

\lstinline|isOwner| checks if the current replica is the candidate with the most votes, and if that candidate has a majority of the votes (assuming a fixed set of voters).

\lstinline|request| votes for the current replica itself if there are no existing votes. 
Note that this implies that it is necessary to retry requesting -- as opposed to the Token algorithm.

\lstinline|upkeep| either (1) transitions to the next epoch if a majority is impossible, (2) does nothing if the replica already voted, or (3) votes for the leading candidate.

\lstinline|release|, if called by the current owner, increases the epoch counter, thus starting a new round of votes.

\subsection{From Protocols to Applications}
\label{protocols to application}

These examples illustrate how to encode mutual exclusion as a data type. 
We have also implemented a protocol inspired by Raft~\cite{ongaroSearchUnderstandable2014} as 
an ARDT, using leader election and then multiple voting rounds based on proposals by that leader.

However, the goal of our endeavour is not to reinvent mutual exclusion or coordination protocols, 
rather to make them available as a composable abstraction in a program.
The discussion so far implies that the LoRe runtime will use one of the mutual-exclusion implementations to coordinate operations when necessary, but we could also re-use components of existing protocols 
as part of the application logic.
For illustration, consider the \code{Excl[T]} type below. 
Given a type \code{T}, it protects values of \code{T} through the Token algorithm; 
the \code{transform} method may be used to change the protected value only if the current replica owns the token.%
\footnote{The \code{Excl} type would need further methods to handle the request/release/upkeep parts of the protocol.}

\begin{lstlisting}[language=lore]
case class Excl[T](token: Token, v: T):
  def transform(f: T => T) =
    if token.isOwner then f(v)
    else unchanged
\end{lstlisting}

This allows a developer to make use of the token algorithm within some larger data type. 
For example, if the application has many resources (many documents in a document management system, or many paragraphs in a single document), then we could specify that each resource is individually mutually exclusive (variant A below), 
or that all resources are protected together (variant B below). By making \code{Excl} parametric over the algorithm, 
we could even choose a different algorithm for each individual resource.

\begin{lstlisting}[language=lore]
val resourcesA: Map[Id, Excl[Resource]]
val resourcesB: Excl[Map[Id, Resource]]
\end{lstlisting}

To recap, our approach allows developers to essentially implement fine-grained application-specific consistency mechanisms 
without having to know the details of the inner data types. 
Notably, the system automatically derives the required lattice instances for all shown data types.

\section{Discussion}\label{discussion}

The ARDT-based implementation of mutual-exclusion protocols offers several benefits, because it allows separation of concerns, while ensuring correctness at the level of the data model of the application.

\paragraph{Composability}

ARDTs are algebraic data structures that are commonly used in programming languages: structs, records, tuples.
Thus, ARDTs are first-class values compatible with existing language features such as parametric polymorphism, pattern matching, data structures\footnote{Basic ARDT set and dictionaries (maps) are just the Scala set/map types}, and third-party libraries.
Notably, all the examples we have shown in Section~\ref{strategy-implementation}, 
perform just standard Scala data-structure manipulation that should be immediately familiar to Scala developers -- and none of this is specific to Scala.

Building on such language features makes ARDTs composable themselves.
As a first step, we have seen how this allows to build mutual-exclusion protocols
by re-using data-structures such as the \code{Epoch} or \code{DotSet} data type.
This principle then extends to the protocol implementations themselves:
While the Voting data type from Listing~\ref{votingprotocol} is essentially a single round voting represented as a set of votes, 
we extend it to a multi-round mechanism by storing it in the \code{Epoch} type that handles multiple rounds abstractly.
Similarly, (a small extension of) the \code{Excl} data type from Section~\ref{protocols to application} can be used to abstract over different mutual-exclusion implementations by swapping out that part of the data type.

By implementing locking protocols as ARDTs we separate the information on which we base decisions from 
the complexities related to low-level message passing, such as unreliability and unordered communication.
For instance, in an ARDT protocol for leader election, \enquote{I vote for X} becomes an operation on a data type;
the developer does not have to specify how this information is expressed as a message,
nor to whom the message is sent, let alone thinking about the order between voting and other operations.
As we discuss next, this separation also facilitates support for a broad range of network protocols.

\paragraph{Minimal Network Requirements}

ARDTs are a good baseline for implementing diverse protocols, because they pose minimal requirements to the network environment.
Specifically, to ensure safety they only require message integrity -- i.e., that messages are not modified during transit -- and to ensure progress they require at-least-once delivery.
Everything else can be handled at a higher level of abstraction, specifically, 
may be handled differently by each data type used in the system.

Notably, this allows us to support diverse network scenarios from WebSockets and \enquote{broadcast between browser tabs} to delay tolerant networks where a message may be stored on a medium and physically transported before arriving at its destination.
Alternatively, we could still choose to send all deltas to a central server, ending up with the standard properties of a client/server based application with offline functionality.
While not every application will need the full range of network support, we believe that having this option \emph{by default} is the only road to ensure resilience in a digital society, where \enquote{standard communication software} should keep working even in unexpected circumstances.

\paragraph{Simplified Causality and Consistency Tracking}

In many systems, coordination happens via an external mechanism (such as off-the-shelf lock managers) sending coordination messages independently of the data.
Thus, there is the question of dependencies (e.g., causality and the \emph{happens before} relation) between coordination messages and operations that change the data.
Expressing coordination as part of data types ensures systematic treatment: Whenever a replica sends an update that releases a certain lock, we can be sure that this update also contains the latest updates to the data that was protected by the lock.
In addition, having one locking ARDT per protected resource (which can be different for each resource) 
allows for very fine-grained tracking of causal dependencies.

\section{Outlook}
\label{outlook}

Our current results are a first step for future work that progresses in two directions. The first is about how to increase flexibility and convenience by expanding the individual ARDTs we currently have into a library of software components to design coordination protocols.
The second is about expanding the verification infrastructure to ensure that the additional flexibility is not invalidated by the impossibility to ensure correctness of the resulting system.

\paragraph{Libraries of Components for Coordination}
\label{locking discussion}
\label{trade-offs}

Distributed systems have trade-offs and assumptions that are interwoven with the design of an application.
Thus, choosing a fitting protocol is not only a matter of the system or the network properties.
For example, a personal note-taking application that synchronizes between the user's devices
may simply assume a fixed number of known devices,
while a chat or microblogging application likely needs to allow participants to join or leave regularly.
The properties that need to be reflected in the application design include:

\emph{Topology:} How many replicas are there? How often does the number of replicas change? Do we allow new replicas to join the network or do we even assume a fixed number?

\emph{Fault-Tolerance:} How reliable are the network connections? How long and often are replicas unreachable?

\emph{Delay-Tolerance:} How important are timely updates? How resilient is the system against stale data?

A particular mutual exclusion protocol presents a single point in the design space on multiple of these dimensions. The example protocols from Section~\ref{strategy-implementation} illustrate this:
The token-passing protocol can deal with arbitrarily changing topologies, but the token is easily \enquote{lost}, causing problems with fault-tolerance in certain scenarios.
The voting protocol improves on resilience against vanishing nodes, but assumes a fixed number of replicas, and requires communication with a majority of those replicas.
More advanced protocols like Raft~\cite{ongaroSearchUnderstandable2014} and Paxos~\cite{lamportParttimeParliament1998, kraskaMDCCMultidata2013, lamportGeneralizedConsensus2005} improve these trade-offs in some dimensions compared to our examples, but they are still bound by fundamental limitations and trade-offs that exist in distributed systems.
Furthermore, \citet{bailisPotentialDangers2012} note that even assuming causal consistency globally can lead to scalability issues, because it leads to causal dependencies that are not required by the application semantics.

Overall, we (1) do need the consistency mechanism to be co-designed with the application, we (2) should be able to make different trade-offs for different parts of the application, and (3) have those parts compose without interfering with one another.
Moreover, (4) developers should not be required to reinvent well-established protocols and systems such as Raft, Paxos, or other systems that already fit their requirements.
We do believe that ARDTs are well-equipped to tackle these challenges.
Specifically, we envision a library of ARDTs that implement or interact with existing mechanisms for mutual exclusion, and other strategies for consistency. This allows reuse (4) and different trade-offs for different parts (2), while interoperability (3) is ensured by expressing everything as an ARDT.  Additionally, we see potential in having smaller components as part of the library, such as voting and epochs. These should enable fine-grained integration of trade-offs into the application design (1).
However, it is still an open question how composition of smaller components would lead to consistent results. We look at a possible solution to this next.

\paragraph{Automatic Verification and Protocol Selection}

Our initial motivation for investigating mutual exclusion protocols was as part of the LoRe verification system, which is currently able to check which invariants may become invalidated due to concurrent operations, and insert the necessary coordination automatically.
In future work, we will consider extensions of this verification infrastructure that integrate with the approach discussed so far.

First, we will use LoRe's existing infrastructure to verify the mutual exclusion of protocols defined as ARDTs, such as the examples in Section~\ref{strategy-implementation}, to guarantee using them is correct.
While formal verification of such correctness is notoriously hard, there are promising results in related work that verifying Raft is simplified when replacing message passing with replicated state~\cite{lewchenkoDistributedConsensus2023}. The reason is similar to the correctness proof of the example in Section~\ref{strategy-implementation}, which abstracts away many of the ordering concerns of message passing systems, and instead reasons about possible sets of values.

As a second step, we aim to extend LoRe's verification to assists with the selection of protocols.
To this end, we envision extending the programming model with annotations beyond safety invariants that enable developers to express systems/deployment constraints.
By leveraging these annotations, we envision it feasible for automated reasoning tools
to choose a fitting protocol or to support the programmer with a pre-selection of fitting choices.

Finally, we hope that future extensions of the verification mechanism in LoRe will be able to take custom consistency mechanisms into account. Specifically, instead of the multiple steps of designing a mutual-exclusion mechanism, proving it correct, and then selecting the mechanism based on certain constraints, we want LoRe to directly prove the invariants while taking the specific ARDTs into account. This would enable very fine-grained protocols that do guarantee application-specific invariants, without needing to adhere to any generic correctness or mutual-exclusion guarantees.

\section{Related Work}
\label{related-work}

Prior work~\cite{lewchenkoDistributedConsensus2023, v.gleissenthallPretendSynchrony2019,farzanSoundSequentialization2022, honoreAdoreAtomic2022} has suggested several techniques to reduce the complexity of message-passing when reasoning about distributed protocols, with the approach by~\citet{lewchenkoDistributedConsensus2023} even directly relying on
modelling consensus protocols as replicated state.
This line of work aims at simplifying the verification of specific protocols (in particular of consensus protocols).
It provides supportive evidence for our goal to avoid message passing in modelling complex protocols.
In contrast, our focus is on the design of a principled approach to implement application-specific consistency protocols, not verify existing ones.

Mixed-consistency has also been widely studied:
\citet{Lewchenko2019} \emph{(Carol)} use an existing replicated data store (IPFS) and every critical operation has to be approved by every other replica in the system.
Similarly, \citet{houshmand2019} \emph{(Hamsaz)} statically analyse conflicts and coordinate them either with a total order broadcast protocol based on consensus or through a locking protocol.
\citet{Alvaro2014} \emph{(Blazes)} propose a program-analysis framework which operates on distributed dataflows.
Based on annotations, the system infers the program's coordination requirements and selects one of two coordination strategies:
\emph{Ordering} enforces a global sequential ordering of critical operations through a Zookeeper~\cite{huntZooKeeperWaitfree2010} instance.
\emph{Sealing} tries to limit coordination to a partition of replicas which actually produce updates that are relevant for the critical operation in question. If such a partition is available, the replicas belonging to the partition participate in a voting protocol to achieve consensus over the current state.
\citet{kohlerRethinkingSafeConsistency2020} \emph{(ConSysT)} offer an information-flow type system,
which distinguishes between \emph{low} and \emph{high} availability objects.
They use centralized master replicas to synchronize data and to manage both levels of availability.
For low availability objects, the master hands out locks to replicas, which prevent concurrent modifications.
\citet{Balegas2015} \emph{(Indigo)} propose a database middleware that coordinates critical operations through multi-level locks.
They offer the locking levels \emph{shared forbid}, \emph{shared allow}, and \emph{exclusive allow},
which allows or forbids replicas the execution of certain operations.
In addition, they offer escrow reservations where certain operations create rights and other operations consume rights.
As long as a replica holds the required rights, it can execute critical operations without coordination.
\emph{IPA}~\cite{Balegas2018} employs automatic compensations and rollbacks in case an invariant violation is detected.
\citet{deporre2021} \emph{(ECROs)} try to reorder non-commutative operations to repair invariant violations.
In case no safe reordering can be found, they resort to coordination.
Gallifrey by \citet{Milano2019} offers mixed-consistency through branching and merging: Non-commutative operations create new branches which can be merged synchronously (requiring coordination) or asynchronously (requiring developer-specified compensations in case of conflicts).

\balance

\bibliographystyle{ACM-Reference-Format}
\bibliography{bibliography}

\appendix
\begin{lstlisting}[float=*t,caption={Code for the Voting protocol.}, label=votingprotocol]
case class Voting(rounds: Epoch[DotSet[Vote]]):

  def @isOwner@(using ReplicaId): Boolean =
    val (id, count) = leadingCount
    id == replicaId && count >= Voting.threshold

  def @request@(using ReplicaId, Dots): Voting =
    if !rounds.value.isEmpty then Voting.unchanged
    else voteFor(replicaId)

  def @release@(using ReplicaId): Voting =
    if !isOwner
    then Voting.unchanged
    else Voting(Epoch(rounds.counter + 1, DotSet.empty))

  def @upkeep@(using ReplicaId, Dots): Voting =
    val (id, count) = leadingCount
    if checkIfMajorityPossible(count)
    then voteFor(id)
    else forceRelease

  def forceRelease(using ReplicaId): Voting =
    Voting(Epoch(rounds.counter + 1, DotSet.empty))

  def voteFor(uid: Uid)(using ReplicaId, Dots): Voting =
    if rounds.value.elements.exists { case Vote(id, _) => id == replicaId }
    then Voting.unchanged // already voted!
    else
      val newVote = rounds.value.addElem(Vote(uid, replicaId))
      newVote.map(rs => Voting(rounds.write(rs)))

  def checkIfMajorityPossible(count: Int): Boolean =
    val totalVotes     = rounds.value.elements.size
    val remainingVotes = Voting.participants - totalVotes
    (count + remainingVotes) > Voting.threshold

  def leadingCount(using ReplicaId): (Uid, Int) =
    val votes: Set[Vote] = rounds.value.elements
    votes.groupBy(_.owner).map((o, elems) => (o, elems.size)).maxBy((o, size) => size)

object Voting:

  val unchanged: Voting = Voting(Epoch.empty)

  given Lattice[Voting] = Lattice.derived

  val participants = 5
  val threshold    = (participants / 2) + 1
\end{lstlisting}

\end{document}